\documentclass[pdflatex,sn-mathphys]{sn-jnl}
\usepackage[normalem]{ulem}
\usepackage{mathtools}

\begin{document}

\title[Sparse RFF IP for HEAs]{Sparse random Fourier features based interatomic potentials for high entropy alloys}
\author[1]{\fnm{Gurjot} \sur{Dhaliwal}}\email{gurjot.dhaliwal@utoronto.ca}

\author[2]{\fnm{Abu} \sur{Anand}}\email{abu.anand@mail.utoronto.ca}

\author*[3]{\fnm{Prasanth B.} \sur{Nair}}\email{prasanth.nair@utoronto.ca}
\author*[1,2]{\fnm{Chandra Veer} \sur{Singh}}\email{chandraveer.singh@utoronto.ca}

\affil*[1]{\orgdiv{Department of Mechanical and Industrial Engineering}, \orgname{University of Toronto}, \orgaddress{\street{5 King's College Rd.}, \city{Toronto}, \postcode{M5S 3G8}, \state{Ontario}, \country{Canada}}}

\affil[2]{\orgdiv{Department of Materials Science and Engineering}, \orgname{University of Toronto}, \orgaddress{\street{184 College St.}, \city{Toronto}, \postcode{M5S 3E4}, \state{Ontario}, \country{Canada}}}

\affil[3]{\orgdiv{Institute for Aerospace Studies}, \orgname{University of Toronto}, \orgaddress{\street{4925 Dufferin St.}, \city{Toronto}, \postcode{M3H 5T6}, \state{Ontario}, \country{Canada}}}

\abstract{Computational modelling of high entropy alloys (HEA) is challenging given scalability issues of Density functional theory (DFT) and non-availability of Interatomic potentials (IP) for molecular dynamics simulations (MD). This work presents a computationally efficient IP for modelling complex elemental interactions present in HEAs. The proposed random features-based IP can accurately model melting behaviour along with various process-related defects. The disordering of atoms during the melting process was simulated. Predicted atomic forces are within 0.08 $eV/\AA$ of corresponding DFT forces. MD simulations predictions of mechanical and thermal properties are within 7\% of the DFT values. High-temperature self-diffusion in the alloy system was investigated using the IP. A novel sparse model is also proposed which reduces the computational cost by 94\% without compromising on the force prediction accuracy.}

\maketitle
\section{Introduction}\label{sec1}
Material discovery has led to significant improvements in the performance and reliability of complex engineering systems, especially for high-temperature applications. A novel class of alloys known as High Entropy alloys (HEAs) has shown excellent properties at extreme temperature conditions \cite{gd1}. HEAs are single-phase alloys of four or more principal elements usually in equiatomic or near equiatomic ratios with excellent structural and functional properties \cite{gd4,gd5}. 

Given the number of possible combinations of the principal elements, it is not feasible to perform experiments on every possible alloy. As such, theoretical and computational methods are used to predict the corresponding phase stability and properties \cite{anand2023recent,gd2,gd3}. The compositional nature of these alloys presents significant computational challenges when using first principle methods such as density functional theory (DFT) \cite{gd3}. An alternative is to use classical Molecular Dynamics (MD) simulations, which provide better computational efficiency than DFT but are limited by the choice of interatomic potential (IP) \cite{gd6}. It is to be noted that the underlying IP must provide accurate potential energy surface for (i) individual elements (more than three in HEAs) (ii) binary interactions of these elements and (iii) complex interactions of HEA as a whole. Most importantly, all these interactions need to be modeled for a range of temperatures and material defects which makes identification of accurate IP a computationally prohibitive task.

Classical IPs based on embedded atom method theory have been developed for HEAs with 4-5 principal components \cite{gd7,gd8,gd9,gd10,gd11,gd12,gd13,gd14,gd15,gd16}. These potentials are used to study a range of mechanical properties and the underlying fracture mechanisms. In \cite{gd8,gd9}, the authors highlighted the importance of lattice distortions and its effects on the mechanical properties of FeNiCuCoCr and FeNiCuCoAl systems. Delugi etal \cite{gd10} used the Embedded atom model (EAM) potential to study the radiation damage response of FeNiCuCoCr alloy. Recently, Huang et al proposed a Modified embedded atom model (MEAM) based potential for the HfNbTaZr system \cite{gd11}. MD simulations performed using these potentials highlighted the importance of chemical short-range ordering on the final quantities of interest. Apart from the successes of the EAM and MEAM potentials, the authors have cautioned against the use of classical IPs and highlighted the limitations of these potentials. The goal of developing EAM potentials in \cite{gd8,gd9} was to study the trends in material properties as affected by the parameters. The authors clearly stated the limitations of the functional form of EAM in describing interactions present in HEA, in particular the underlying elemental interactions. Another limitation was the description of pair interactions which were assumed to be the average of the elemental potentials. 

An alternative to classical IPs is to use machine learning (ML) methods for IP development \cite{gd17,gd18}. Some of the popular IP choices are based on Neural networks \cite{gd20}, Gaussian processes \cite{gd21}, linear models \cite{gd23,gd24,gd29} and kernel regression models. A review showing the applicability and limitations of ML-based IPs can be found here \cite{gd19, gd22}. Apart from their huge success in developing IP for binary and ternary alloys, most of the ML-based models are limited by computational expense when applied to model HEAs. Neural networks are limited by the data requirements and long training time while kernel-based methods are limited by memory requirements. To ease the computational burden during training and prediction stages, tabulated versions based on spline approximation of Gaussian Approximation Potentials (GAP) have been proposed \cite{gd25,gd26}. The developed potentials were used to study defect evolution in MoNbTaVW \cite{gd26,gd27}. Though computationally faster than kernel methods, the accuracy of this potential is limited by the step size of the spline approximations. 

Due to the computational burden associated with neural networks and GAP, most of the recent work in this field has been focused on generalized linear models. In \cite{gd28}, the authors generated Smooth Overlap of Atomic Positions (SOAP) features and fitted a linear model to predict the energy/forces for NbMoTaW. The major drawback of this potential is the computational cost associated with generating the SOAP features. For larger systems, the feature generation process can be memory intensive as well. Another popular choice is based on the low-rank representation of the potential energy tensor \cite{gd30,gd31,gd32}. This potential is extensively used to understand phase transitions and the effects of local disorders on the mechanical properties of various HEAs. The potential considers nearest neighbours and is well suited for equi-phase elements. For systems, as considered in this paper, with elements having a mix of phases (BCC and FCC), this potential might not be suitable. Another popular choice is Moment tensor potential (MTP) which has been used to compute vibrational free energies and mechanical properties of HEAs \cite{gd33,gd34,gd35}.

This work presents a novel generalized linear model that leverages random Fourier features to efficiently approximate the IP of HEAs. The presented model can accurately model the complex elemental, binary and quaternary interactions in the FeNiCuCoCr (FNCCC) system. As compared to the previously published EAM potentials for this system, we considered FCC phases for Ni, Co and Cu while BCC phases for Fe and Cr. Further, the model only requires the computation of simple features such as two-body and three-body descriptors, as compared to the computation of complex features such as SOAP and moment tensor descriptors. The random Fourier feature-based IP was tested for a range of MD simulations focusing on structural, mechanical, and thermal properties. Simulated property values are within 7\% of the corresponding DFT values. We further reduced the computational burden by utilizing a sparse Bayesian learning approach. Sparse models reduced the number of original features by 94\% while maintaining the same level of accuracy for energy/force prediction. MD simulations further confirmed that results from sparse models for FNCCC are within 11\% of corresponding DFT calculations. 

\section{Methodology}\label{methodologySec}

\subsection{Kernel modeling}
The material system considered in this study has five distinct elements i.e. Fe, Ni, Cu, Co, Cr. Following the terminology presented in \cite{gdNPJRFF}, we consider $L$ atomic configurations. We assume that the $l^{th}$ configuration has $N_l^{a}$ atoms of type $a$, where $a \in \mathcal{S}\coloneqq\{Fe, Ni, Cu, Co, Cr\}$. The energy of the $l^{th}$ atomic configuration calculated using DFT is denoted by $E^{l*}$. DFT calculated force on the $i^{th}$ atom in the $l^{th}$ configuration is denoted by $F_{u,i}^{l,a,*}$, where $u$ is the direction of the force and $a$ is the atom type.

Each atom is described by a set of rotational, translational and permutation invariant descriptors. The popular choice of descriptor includes the distance between two atoms, bispectrum coefficients, Behler's symmetry functions and SOAP \cite{gd22}. Each descriptor varies in terms of accurate representation of the atomic environment and computational complexity. We define an atomic descriptor corresponding to atom $i$ of type $a$ as $q_{i}^{a}$. The goal is to learn local atomic energy and forces, given the descriptors.

In the present work, the IP model is based on the kernel representation of the local atomic energy.  The local atomic energy of an atom belonging to configuration $l$ and described by the descriptor $q_{i}^{b}$ can be written as 

\begin{equation}
    E^{l}(q_{i}^{b}) = \sum_{t=1}^{L}w_{t}\sum_{a\in \mathcal{S}}\sum_{j=1}^{N_{t}^{a}}K^{ab}(q_{i}^{b},q_{j}^{a}), 
\end{equation}
where $K^{ab}(q_{i}^{b},q_{j}^{a})$ denotes the kernel or similarity matrix between the two atoms, of type $a$ and $b$ that are described by the descriptors $q_{i}^{b},q_{j}^{a}$ respectively. Energy contribution from each atomic configuration is scaled by the weights $w_{t}$.

In general, values for energy and forces are provided in the training dataset based on DFT calculations. Thus, parameter estimation requires an inversion of a $N\times N$ matrix which costs $\mathcal{O}(N^3)$, where $N$ is the total number of energy and force values. For a $D$ dimensional descriptor $q_{i}^{b}$, the prediction cost is of the order $\mathcal{O}(ND)$. For the IP training problems, $N$ can easily reach $10^6 - 10^7$ and $D$ can be easily of the order 2. This is a serious computational bottleneck and limits the use of kernel-based IPs for HEA applications. Another limitation of the kernel model is the memory requirement during the prediction stage. The entire training dataset needs to be loaded in Random Access Memory (RAM) during the prediction stage which can limit the scalability of the simulation cell.

Considering only two-body and three-body descriptors, we need to model 25 two-body interactions and 75 three-body interactions for FNCCC alloy. Even for these simple descriptors, we need at least 100 separate kernels that model these interactions accurately. Our application requires modelling various process conditions for elemental and FNCCC alloys. This easily results in a training dataset with more than 1.3 million atomic configurations. Given the computational cost and memory requirements, it will be challenging to develop a kernel-based IP using a dataset of this scale.

\subsection{Random features}
To alleviate the computational burden, we used efficient kernel approximation methods based on random features. The idea of random features for machine learning-related applications was first introduced in \cite{gd36}. Random features proposed in \cite{gd36} assume that the underlying kernel is stationary. Its application to model IP was first explored in \cite{gdNPJRFF}. The work in \cite{gdNPJRFF} developed random features based IP for single element systems and demonstrates transfer-ability across three distinct material classes. In general, accurate energy modelling using SOAP features requires non-stationary kernels such as higher-order polynomial kernels. The work in \cite{gdNPJRFF} presented efficient generalized linear IP models for stationary as well as non-stationary kernels.

Random features approximate the underlying kernel as
\begin{equation}
    K^{ab}(q_{i}^{a},q_{t}^{b}) \approx z(q_{i}^{a})^{T}z(q_{t}^{b}),
\end{equation}
where $z(q_{i}^{a})$ denotes the random features associated with the atom defined by the descriptor $q_{i}^{a}$. For stationary kernels such as the squared exponential kernel, random features can be written as  
\begin{equation} \label{RFFEqn}
z(q_{i}^{a}) = \sqrt{\frac{2}{M}}\{\cos(\omega_{1}q_{i}^{a}),\sin(\omega_{1}q_{i}^{a}), \cdots,  \cos(\omega_{M}q_{i}^{a}),\sin(\omega_{M}q_{i}^{a})\},
\end{equation}
where $\omega$ is a random vector as drawn from a probability distribution $p(\omega)$ and $M$ is the number of random features \cite{gd36}. The above approximation is based on Bochner's theorem which states that a positive definite stationary kernel can be written as a Fourier transform of a particular measure, $p(\omega)$ \cite{gd36}. Taking the Monte-Carlo estimate of the resulting Fourier transformation leads to equation (\ref{RFFEqn}). The sampled $\omega$ can be constrained to be orthogonal, which can result in a reduced basis of random features without losing accuracy in MD simulation prediction \cite{gdNPJRFF}. As compared to the standard kernel models, the computational complexity associated with parameter estimation and prediction scale is $\mathcal{O}(M^2N)$ and $\mathcal{O}(M)$, respectively.  

\subsection{Energy model}
We modelled the local atomic energy using a generalized linear model based on random Fourier features. For an atom in the $l^{th}$ training configuration and described by the descriptor $q_{i}^{a}$, we consider a model for the local atomic energy of the form 
\begin{equation} \label{eqn4}
    E^{l}(q_{i}^{a}) = \sum_{m=1}^{2M}\alpha_{m}z_{m}(q_{i}^{a}),
\end{equation}
where $z_{m}(q_{i}^{a})$ is the $m^{th}$ component of the vector defined in (\ref{RFFEqn}) and $\alpha_m$ is the corresponding weight. The force on this atom, in direction $u$, is given by the following equation
\begin{equation}
    F^{l}_{u}(q_{i}^{a}) = -\frac{\partial}{\partial r_{u}}(E^{l}(q_{i}^{a})).
\end{equation}

A suitable cutoff function can also be applied with the above energy model to cancel the atomic interactions falling beyond a cutoff radius. 
\subsection{Parameter estimation}
The parameters $\alpha$ are obtained by minimizing the following loss function

\begin{eqnarray} \label{lossFunc}
    loss(\alpha) &=& \sum_{l=1}^{L}\left(\left(\sum_{a \in \mathcal{S}} \sum_{i=1}^{N_{l}^{a}}E^{l}(q_{i}^{a})\right) - E^{l*}\right)^{2} \nonumber \\&+& \lambda_1\sum_{l=1}^{L}\sum_{a \in \mathcal{S}}\sum_{i=1}^{N_{l}^{a}}\sum_{u=1}^{3}\left(F^{l}_{u}(q_{i}^{a}) - F^{l,a,*}_{u,i}\right)^{2} + \lambda_2 \|\alpha \|^{2}_{2},
\end{eqnarray}
where quantities marked with an asterisk are DFT-calculated quantities. Errors in force prediction are weighted by the scalar $\lambda_1$ and the regularization parameter is denoted by $\lambda_2$. Both of these hyper-parameters can be determined through cross-validation. 

\subsection{Sparse Bayesian learning of IPs}\label{RVMSec}
Random features-based potentials can circumvent the limitations of kernel models. To reduce the number of terms in the generalized linear model, various sparse approximations can be used. This can be beneficial for reducing the run-time complexity of MD simulations using the learned IP model. In this work, we explored an efficient sparse Bayesian learning method known as the Relevance Vector Machine (RVM) \cite{gd37}. The main idea behind RVM is to select the most informative basis functions by using an appropriate prior on the weights in (\ref{eqn4}). RVM assumes this prior to be a Gaussian with a separate noise hyper-parameter for each weight $\alpha_i$
\begin{equation}\label{priorRVM}
    p(\alpha \vert \beta) = \prod_{i=1}^{2M} \mathcal{N}(\alpha_i\vert 0,\beta_i^{-1}),
\end{equation}
where $\beta_i$ is the precision for each IP parameter. From the Bayesian perspective, minimization of (\ref{lossFunc}) yields the maximum a posteriori estimate of the weights when a Gaussian prior with the same precision is chosen for each $\alpha_i$. In contrast, RVM considers separate noise parameters. It can be easily shown that the prior in equation (\ref{priorRVM}) results in a regularization term of the form $\sum_{i=1}^{2M} \log{\alpha_i}$ \cite{gd38}.

Training of RVM involves identifying optimized values of $\alpha$ along with noise and weight hyper-parameters. The optimization involves maximizing the type-II marginal likelihood. In practice, optimizing the marginal likelihood leads to many individual $\beta_i's$ tending to infinity and thus corresponding $\alpha_i$ becomes close to zero. The marginal likelihood can be maximized using gradient descent or expectation-maximization algorithm. A faster version based on sequential estimation is provided in \cite{gd39}. RVM has been used in various diverse prediction tasks, but its use in IP modelling has not been explored yet.

\section{Data generation}
\subsection{DFT Simulations}
Training data was generated using plane wave Density Functional Theory (DFT) calculations conducted using the Vienna Ab initio Simulation Package (VASP)\cite{kresse1993ab,kresse1996phys}. Interactions between the valence electrons and the ionic core were described by generalized gradient approximation (GGA) in the Perdew-Burke-Ernzerhof (PBE) formulation along with projector augmented wave (PAW) pseudopotentials \cite{blochl1994projector,kresse1999ultrasoft,perdew1996generalized}. Brillouin zone sampling was done using the \textit{k}-point mesh generated using the Monkhorst-Pack scheme \cite{monkhorst1976special}. Spin polarization was considered for all DFT calculations.  
\subsection{Training Data}
Data required for training was generated in gradual steps. We started with a database of elementary interactions. For each element, we generated the ground state of the respective phases as well as the corresponding strained geometries. To accurately capture the melting behaviour of each element, we performed AIMD simulations from 300K to 2500K for each element and added sampled snapshots into the training data. 

For the FNCCC alloy, we generated a single random structure and relaxed it using DFT. The entire trajectory corresponding to this relaxation was included in the training set. In the next step, we added strained geometries of this alloy. A single point vacancy, corresponding to each element, was introduced in the pristine structure. The resulting defective structures were relaxed using DFT. Snapshots from the optimized trajectory were included in the training. Finally, the melting behaviour of the FNCCC alloy was simulated using AIMD simulations performed in the temperature range of 300K – 2500K. Samples from the melting trajectory were included in the training set. For each training dataset, Table \ref{Table:TrainingData} below shows the number of structures, the number of atoms corresponding to each element, along the total number of atoms in the respective dataset.

\begin{table}[!ht]
\begin{tabular}{c c c c c c c c}
Dataset & $N_{s}$ & $N_{atoms}$ & $N_{Fe}$ & $N_{Ni}$ & $N_{Cu}$ & $N_{Co}$ &$N_{Cr}$ \\
\hline
Elemental Ground & 83 &	9564 &	2208 &	1872 &	1656 &	1656 &	2172\\
Elemental Strained &	1586 &	181408 &	32,384 &	38,880 &	38,880	& 38,880 &	32,384\\
Elemental Melting &	774	& 89712& 	19584 &	16524 &	17496 &	16524 &	19584\\
FNCCC Strained &	822 &	88776 &	19728 &	17262 &	17262 &	17262 &	17262\\
FNCCC Vacancy &	120 &	12840 &	2860 &	2500 &	2480 &	2500 &	2500\\
FNCCC Melting &	264 &	28512 &	6336 &	5544 &	5544 &	5544 &	5544\\

\end{tabular}
\caption{\label{Table:TrainingData} Details about the training data are shown in this table. The second column in the table denotes the number of structures in the respective dataset. The total number of atoms in the dataset is shown in the third column followed by the number of atoms corresponding to each element in the dataset.}
\end{table}

\section{Results}\label{resSec}
Using DFT data as mentioned in the previous section, we trained a potential based on random Fourier features for the FNCCC alloy. We used orthogonal random numbers to generate random Fourier features. The potential considered 25 two-body interactions and 75 three-body interactions. Each two-body interaction was approximated using 60 random features and each three-body interaction was approximated using 180 random features. The number of parameters for this model is 15000, which were estimated by minimizing the objective function in (\ref{lossFunc}), using efficient numerical schemes. This potential will be referred to as O-RFF in the rest of the paper.
\begin{figure}[!ht]
\centering
\includegraphics[trim=36 50 70 70,clip,scale=0.4]{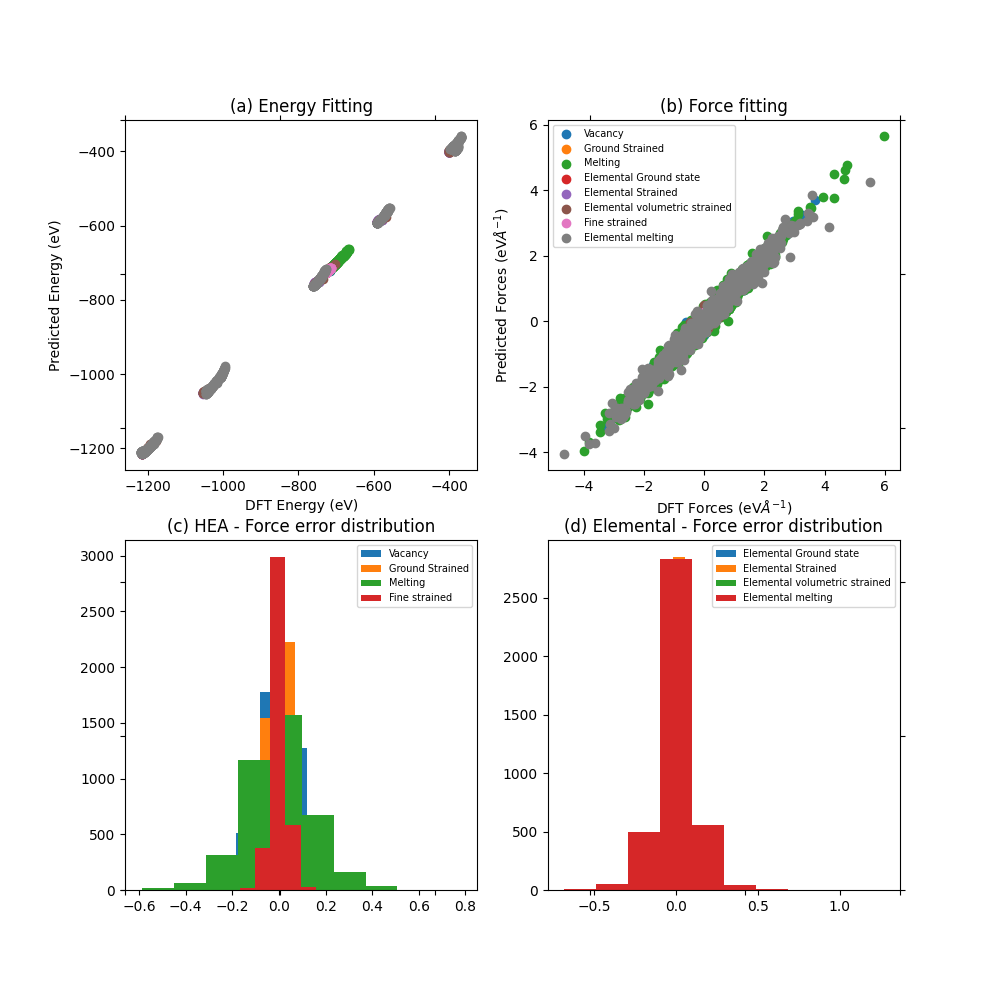}

\caption{\label{f:fig1} Performance comparison of O-RFF IP to DFT calculations of (a) Energy and (b) Forces. The error distribution of forces for HEA data is shown in (c) and for the elemental system in (d).}
\end{figure}

Figure \ref{f:fig1} compares DFT calculated energy and forces to the ones predicted through O-RFF. The mean absolute error (MAE) for energy prediction is less than 0.031 eV/atom for all FNCCC and elemental interactions. The corresponding value of MAE for force prediction is in the range 0.03 – 0.11 eV/angstrom. Figure \ref{f:fig1}(c) also shows the distribution of the error in force prediction for FNCCC while the corresponding distribution for the elemental datasets is shown in Figure \ref{f:fig1}(d). From both histograms, it is clear that O-RFF can model the elemental interactions as well as complex interactions of FNCCC alloy with DFT-level accuracy. 

\subsection{Mechanical Properties}\label{mechPropSec}
The validity of O-RFF is assessed by performing a range of MD simulations for each element and comparing them to existing experimental and DFT studies. All MD simulations in this study are performed using the Atomic Simulations Environment package \cite{aseRef}. Elastic constants were computed for each element by fitting a linear relation on stress and strain values. For each element, a $4 \times 4 \times 4$ simulation box was created resulting in 108 atoms for FCC phased elements and 128 atoms for BCC phased elements. Each simulation box was incrementally strained for several discrete finite steps. At each step, the strained simulation box is relaxed by minimizing the maximum force present in the atoms. After relaxation, the stress tensor on each strained configuration is computed. Elastic constants are calculated by applying a linear fit on the corresponding stress and strain values.

\begin{table}[ht]
\resizebox{1.0\textwidth}{!}{
\begin{tabular}{c c c c c c c c}
& Method & C$_{11}$ (GPa) & C$_{12}$ (GPa)& C$_{44}$ (GPa)& Bulk modulus (GPa)& Poisson ratio & Young's modulus (GPa) \\
\hline
Ni & DFT & 246.73 & 158.1 &119.23 &187.64 & 0.391 &123.24 \\
   &O-RFF&261.23 & 176.71 & 110.75 & 204.88 & 0.404 & 118.63\\
&SparseRFF&288.4 & 162.74 & 131.46 & 204.63 & 0.361 & 170.99\\
\hline
Fe & DFT& 268.63 & 139.15 & 102.99 & 182.27 & 0.341 & 173.73 \\
   &O-RFF&268.34 & 149.8 & 120 & 189.31 & 0.358 & 161.01\\
&SparseRFF&278.14 & 150.31&	91.49&	192.92&	0.351 & 172.68\\
\hline
Cu &DFT & 174.06 & 121.29 & 60.35 & 138.88  & 0.411 & 74.44 \\
   &O-RFF&177.34 & 116.66 & 67.92 & 136.89 & 0.397 & 84.76\\
&SparseRFF&174.78&	115.61&	63.86&	135.33&	0.398&	82.73\\
\hline
Co & DFT& 266.7 & 161.87 & 136.19 & 196.81  & 0.378 & 144.42\\
   &O-RFF&277.45 & 190.58 & 137.36 & 219.54 & 0.407 & 122.24\\
&SparseRFF&269.01 & 175.72 & 150.7 & 206.82 & 0.395 & 130.15\\
\hline
Cr &DFT & 488.75 & 156.85 & 94.03 & 267.48  & 0.243 & 412.54 \\
   &O-RFF&503.73 & 161.55 & 85.14 & 275.61 & 0.243 & 425.27 \\
&SparseRFF& 515.29 & 145.83 & 105.01 & 268.98 & 0.221 & 450.96\\
\hline
FNCCC & DFT &196.04 & 131.89 & 113.55 & 153.27 & 0.402&89.95 \\
    &O-RFF& 194.43 &	129.95 &	105.4 &	151.44 &	0.401 & 90.311 \\
&SparseRFF& 175.39 & 137.36 & 106.85 & 150.04 & 0.439 & 54.73 \\
\end{tabular} 
}
\caption{\label{Table:1} Molecular dynamics simulation results for O-RFF potential as compared to DFT calculations. Also, shown are the results for SparseRFF potential as explained in the section \ref{SparseResults}.}
\end{table}

Table \ref{Table:1} shows the fitted elastic constant values for each element as computed by O-RFF and compared to the corresponding DFT values. C$_{11}$ for each element is within 6\% of the corresponding DFT values of each element. Similarly, C$_{12}$ for the elements is within 7\% except for Ni and Co, which show a deviation of 20 GPa and 30 GPa, respectively. The values for C$_{44}$ also lie within 9 GPa of DFT values except for Fe which shows a deviation of 20 GPa. The discrepancy in C$_{12}$ and C$_{44}$ can be handled through hyperparameter tuning and setting higher weights for these configurations in the training dataset. 

The bulk modulus (B) for each element is calculated from the predicted elastic constant values using the following relation $B=  (C_{11}+2C_{12})/3$.
Similarly, we computed the Poisson’s ratio($\nu$) as, $\nu=  1/(1+ \frac{C_{11}}{C_{12}})$.
Using the above values of B and $\nu$, we calculated Young’s modulus (E) as $E=3B(1-2\nu)$.

From Table \ref{Table:1}, bulk modulus for all the elements is within 3\% of the DFT values, except for Ni and Co. The discrepancy in bulk modulus prediction for these two elements can be attributed to an error in $C_{12}$ predictions. Similarly, the variation of Poisson’s ratio for all the elements except Cobalt is within 5\% of the DFT value. The over-prediction in the case of Cobalt can be attributed to the error in $C_{12}$ values. The last column of Table \ref{Table:1} shows the variation in Young’s modulus which is within 15\% of the DFT values.

\begin{figure}[!ht]
\centering
\includegraphics[trim=-1 0.1 0.1 0.1,clip,scale=0.5]{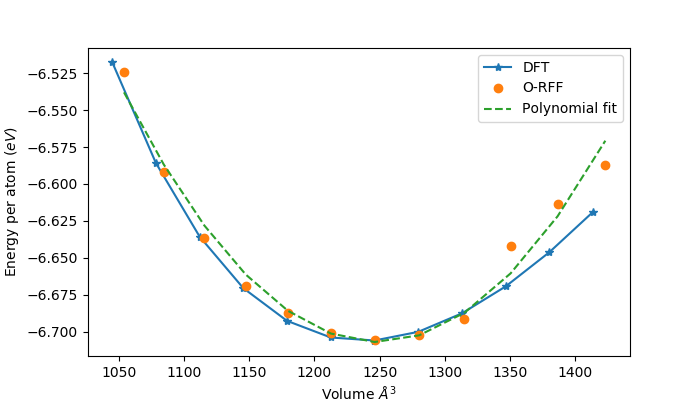}

\caption{\label{f:fig2}
Energy-volume relationship for FNCCC as computed using DFT shown as a solid blue curve. O-RFF predicted energy volume values are shown as green dots. A polynomial fit is performed on the O-RFF predicted energy-volume values and is shown as an orange curve.}
\end{figure}

The mechanical properties of the FNCCC alloy were computed using the simulation methods as explained above. An initial simulation cell of the FCC lattice is created. Each lattice point is filled with an atom from our element list using random selection while maintaining equi-atomic ratios. For the calculation of cohesive energy, lattice constant and elastic constants, a simulation box consisting of 108 atoms was created. Figure \ref{f:fig2} shows the energy-volume relationship for FNCCC alloy. Cohesive energy from this plot comes out to be -6.706 eV/atom and matches the corresponding DFT value

Table \ref{Table:1} compares the mechanical properties of the FNCCC alloy as computed by O-RFF to the ones calculated through DFT. Both $C_{11}$ and $C_{12}$ are within 1\% of the DFT calculated values. O-RFF under-predicted the $C_{44}$ value by 7\%. Quantities such as bulk modulus, Young’s modulus and Poisson’s ratio show excellent agreement with DFT results with each quantity within 1\% of the corresponding DFT value. 

\subsection{Vacancy Formation Energy}
The vacancy formation energy ($E^f_v$) corresponding to an element $a$ can be computed using the following equation

\begin{equation}
    E^{f}_{v} = E^{t*} - E^{0} + \mu_a,
\label{vacEqn}
\end{equation}
where $E^{t*}$ is the energy of the relaxed vacancy system and $E^{0}$ is the energy of the relaxed bulk system with $N$ atoms. Chemical potential for an element $a$ is denoted by $\mu_a$. The number of atoms in the defective cell is one less than in the pristine cell. Apart from energy computation of the defected and pristine cell, detailed calculations are needed to estimate the chemical potential values. Calculation of chemical potentials can be avoided by the statistical methodology as presented in \cite{zhang2021statistical}. The idea is to obtain an average vacancy formation energy ($\langle E^f_v \rangle$) which is estimated by considering $N$ defective structures, each containing the single atom vacancy. From these $N$ defective structures, the average vacancy formation energy can be computed as 

\begin{equation}
    \langle E^{f}_{v}\rangle = \frac{1}{N}\sum_{k=1}^{N}E^{t*}_k - \frac{N-1}{N}E^{0},
\label{avgVacEqn}
\end{equation}
where $E^{t*}_k$ is the energy of $k^{th}$ defective structure. The chemical potential for each element can also be estimated using mean vacancy formation energy as 

\begin{equation}
    \mu_j = E^{0} + \langle E^{f}_{v}\rangle - \frac{1}{N_j}\sum_{k=1}^{N_j}E_k^{t*},
\label{chemPotEqn}
\end{equation}
where $N_j$ is the number of defective structures with a single vacancy of atom type $j$. A detailed derivation of the above equations can be referred to at \cite{zhang2021statistical}. The mean vacancy formation energy is computed on a supercell of 108 atoms. Defected structures are created by removing a single atom from this cell. To obtain an accurate estimation of average vacancy formation energy and chemical potential, 108 simulations corresponding to each atom in the supercell are performed. Mean vacancy formation energy and chemical potentials were estimated using both DFT and MD simulations. For MD simulations, we considered EAM \cite{gd10} and O-RFF potentials. For each method, the defective structure is relaxed using force minimization.


\begin{figure}[!ht]
\centering
\includegraphics[trim=-1 0.1 0.1 0.1,clip,scale=0.4]{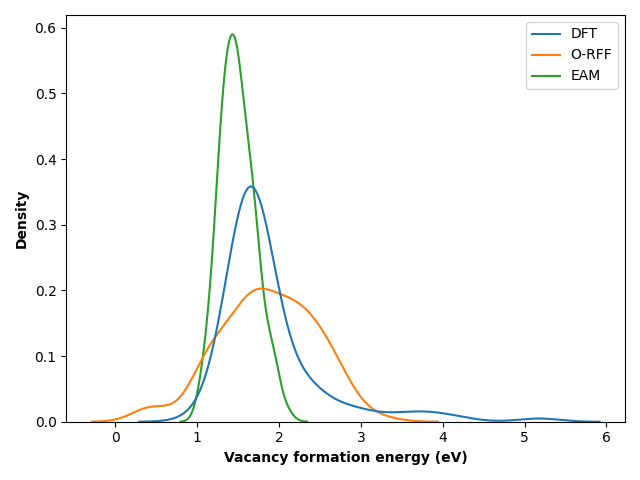}

\caption{\label{f:fig3} Distribution of Vacancy formation energy as computed using DFT and MD simulations. Average vacancy formation energy as computed using O-RFF shows less bias as compared to the one computed using EAM. }
\end{figure}

\begin{table}[!ht]
\centering
\begin{tabular}{c c c c }
        &DFT (eV) & EAM (eV) & O-RFF (eV)\\
\hline
$\langle E^{f}_{v}\rangle$ &1.886 &1.49  &1.859  \\
$\mu_{Ni}$                 &-5.592&-4.435&-5.784 \\
$\mu_{Fe}$                 &-8.192&-4.281&-8.281 \\
$\mu_{Cu}$                 &-3.432&-3.613&-3.679 \\
$\mu_{Co}$                 &-6.982&-4.446&-6.657 \\
$\mu_{Cr}$                 &-9.123&-4.213&-8.907 \\
\end{tabular}
\caption{\label{Chap6:Table:2} Mean vacancy formation energy as predicted by DFT, EAM and O-RFF potential. Chemical potential $\mu_A$ for each element $A$ is also shown here and compared against the DFT obtained values.}
\end{table}

Figure \ref{f:fig3} shows the vacancy formation energy as calculated by O-RFF and EAM potentials and compared against the corresponding values obtained using DFT. Mean vacancy formation energy from the three methods is detailed in table \ref{Chap6:Table:2}. O-RFF predicted mean vacancy formation energy is closer to DFT as compared to EAM potential. Also, the variance in O-RFF predicted energies is slightly higher than the DFT obtained values. Table \ref{Chap6:Table:2} also shows the computed chemical potentials and is compared against the DFT obtained values. The per cent deviation between the DFT obtained and O-RFF computed chemical potentials is less than 0.1\%. Chemical potential calculation shed some light on the applicability of O-RFF for vacancy formation energy. Since FNCCCC is a random solid solution, chemical potential should not be equal to the cohesive energy of the corresponding element, which is accurately captured by O-RFF. On the other hand, EAM provides cohesive energy of each element as chemical potential.

\subsection{Stacking Faults and Dislocation Splitting}
\begin{figure}[!ht]
\centering
\includegraphics[trim={1 15 1 10}, clip, scale=0.56 ]{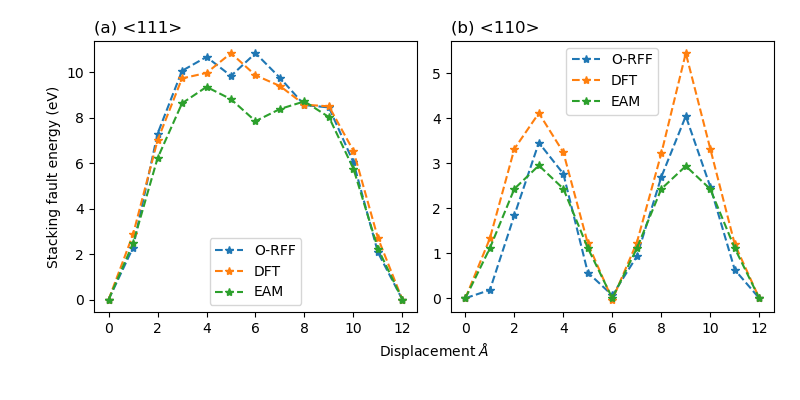}
\caption{\label{f:figSF} Stacking fault energy curves obtained for direction 111 (a) and 110 (b), using DFT, O-RFF and EAM potential.}
\end{figure}

Generalised stacking fault energy curves (GSFE) in the \{112\} plane were simulated along the $<$110$>$ and $<$111$>$ directions for the FNCCC alloy system. A similar methodology as described by Ojha \textit{et al.}\cite{ojha2014twinning} is utilized. FNCCC supercell was systematically sheared and allowed to relax along the Z-direction to eliminate any in-plane forces. Figure \ref{f:figSF} compares the GSFE curves obtained using the O-RFF potential with those calculated using DFT and the EAM potential \cite{EAM_IP}. Maximum absolute error between DFT and O-RFF predicted stacking fault energies is within 1.1 eV. This exhibits that the O-RFF, potential developed in this work, can accurately reproduce the GSFE curves for both directions. The curves showed typical FCC behaviour except for some ruggedness. This could be attributed to the random chemical environment in the FNCCC system. 

\begin{figure}[!ht]
\centering
\includegraphics[scale=0.20 ]{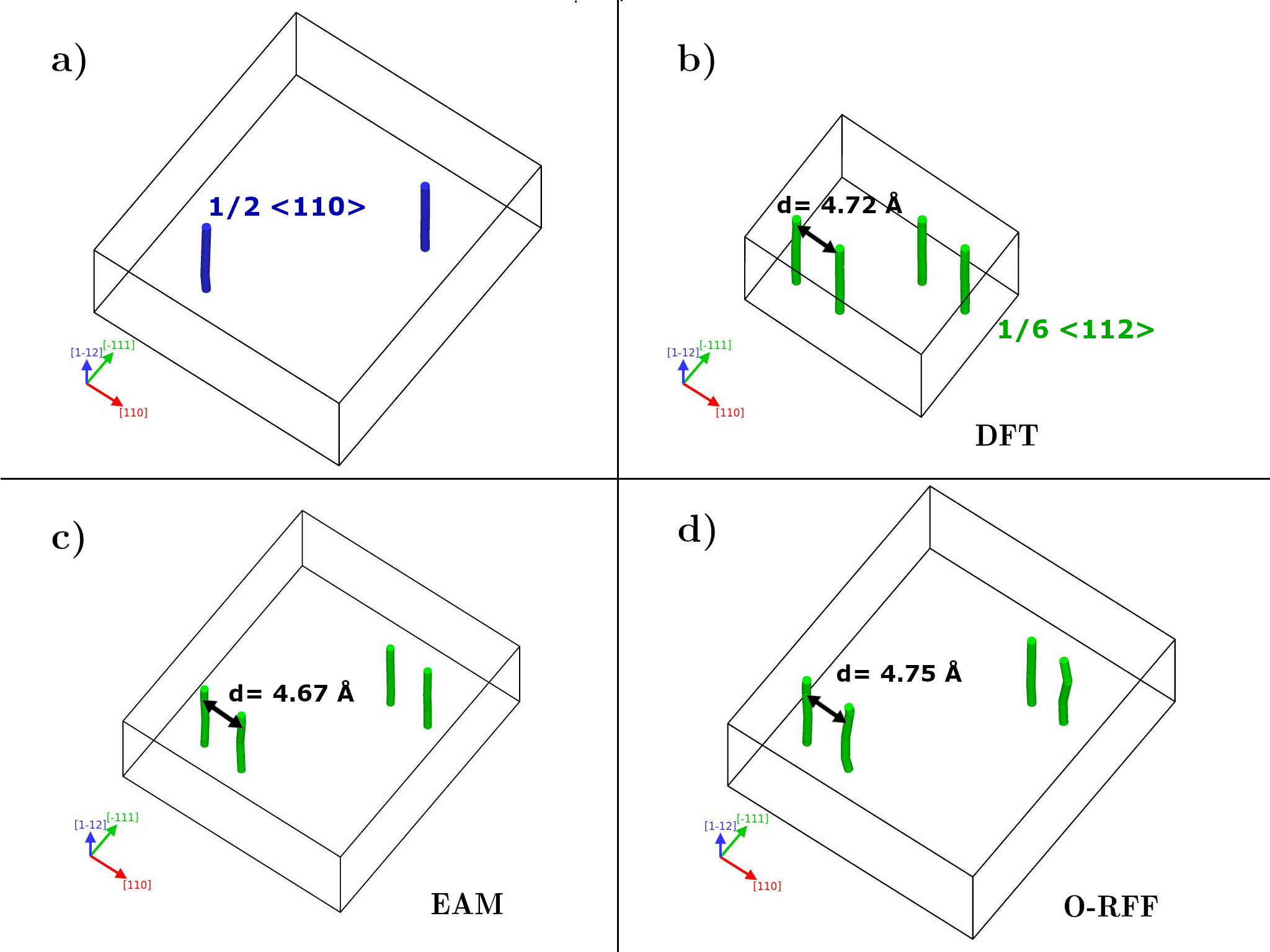}
\caption{\label{f:disloc_split} a) Edge dislocation dipole prior to relaxation with full dislocations (Blue). (b)-(d) Dislocation dipole after relaxation splitting into Schokley partials (Green).}
\end{figure}

Dislocation splitting was investigated in the FNCCC system. An edge dislocation dipole with perfect dislocations ($\frac{1}{2}<110>$ type) was created using Atomsk \cite{hirel2015atomsk}. A dipole geometry was chosen to maintain a zero net Burgers vector. Upon relaxation the perfect dislocations were split into Schokley partials of type $\frac{1}{6}<112>$ with DFT, EAM\cite{EAM_IP} and O-RFF potentials (Figure \ref{f:disloc_split} (b)-(d)). The dissociation distance of the partial dislocations ($d$) was found to be comparable across the different methods.

\subsection{Melting and Diffusion coefficients}
\label{sec:melting}
Using the O-RFF potential, we studied the melting behaviour of the FNCCC alloy. An FCC simulations cell of dimension $21.4\AA \times 21.4\AA \times 21.4\AA$ was constructed, and positions were filled randomly with atoms from the design space. We specified the initial velocity distribution corresponding to 300K and raised the temperature to 2500K in discrete steps of 100K. At each step, the simulation box was equilibrated at the corresponding temperature using an NPT ensemble. The equilibration at each temperature was performed for 1.2 picoseconds using a time step of 4 femtoseconds. The temperature rise from 300K to 2500K was performed in 27.6 picoseconds. 

\begin{figure}[!ht]
\centering
\includegraphics[trim=-1 0.1 0.1 0.1,clip,scale=0.4]{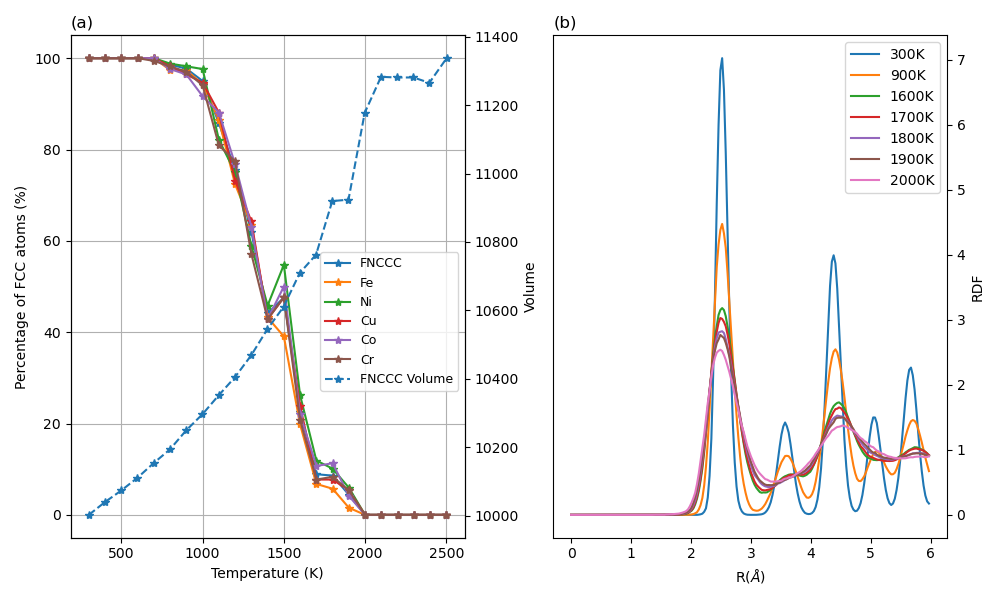}

\caption{\label{f:fig4} Variation of volume with respect to temperature as well as percentage of FCC atoms at each temperature are shown in (b). Radial distribution function plots at various temperatures are shown in (b).}
\end{figure}

Figure \ref{f:fig4}(a) shows the variation of average volume with respect to the average temperature for the resulting trajectory. We also performed common neighbour analysis using the Polyhedral matching template algorithm as implemented in OVITO \cite{ovitoRef}. For the entire simulation, we only observed a change in the FCC phase into other phases. Figure \ref{f:fig4}(a) also shows the percentage of FCC atoms corresponding to each temperature.  As expected, with an increase in temperature, an increase in volume and a decrease in the crystal symmetry are observed. 

The radial distribution function (RDF) of the FNCCC alloy at various temperatures is also shown in Figure \ref{f:fig4}(b). At 300K, there are sharp peaks at the first and second neighbour distances, indicating FCC symmetry. Raising the temperature to 900K reduces the peak height and increases the curve width. From Figure \ref{f:fig4}(a), we can see that this is the temperature where the FCC phase starts converting into other phases. The percentage of FCC atoms continues to decline steadily, and a sharp drop is observed starting at 1500K but not much change in the slope of the temperature-volume curve is observed. Further, another sharp drop is observed between 1600K – 1700K and a slight change in the slope of the temperature-volume curve is also noticed, indicating the melting process has started. At 1700K, only 10\% of the atoms are in FCC state which steadily goes to zero at 2000K. RDF plots are showing similar density beyond 1700K. Also, we noticed a sharp change in the slope in the temperature-volume curve at 1700K. From these observations, we conclude that the melting temperature for FNCCC should lie in the range of 1500 - 2000K. Experimentally, the melting temperature of this alloy is found to be 1500K \cite{meltFNCCC}, which further corroborates our observation regarding the melting temperature.

For a particular temperature of interest, the MSD of the atoms can be extracted from the MD trajectory using the following equation known as the Einstein formula

\begin{equation}
    \label{e:msd}
    MSD = \Bigg \langle \frac{1}{N_{atoms}} \sum_{n=1}^{N_{atoms}} \vert r_{\alpha}^{n} - r_{\alpha}^{n}(t_0) \vert^2 \Bigg \rangle_{t_{0}},
\end{equation}
where $N_{atoms}$ is the number of atoms in the system and $r_{\alpha}^{n}$ are their coordinates which are averaged over time origins \cite{mei1990molecular, maginn2019best}.

\begin{figure}[!ht]
\centering
\includegraphics[width=\textwidth]{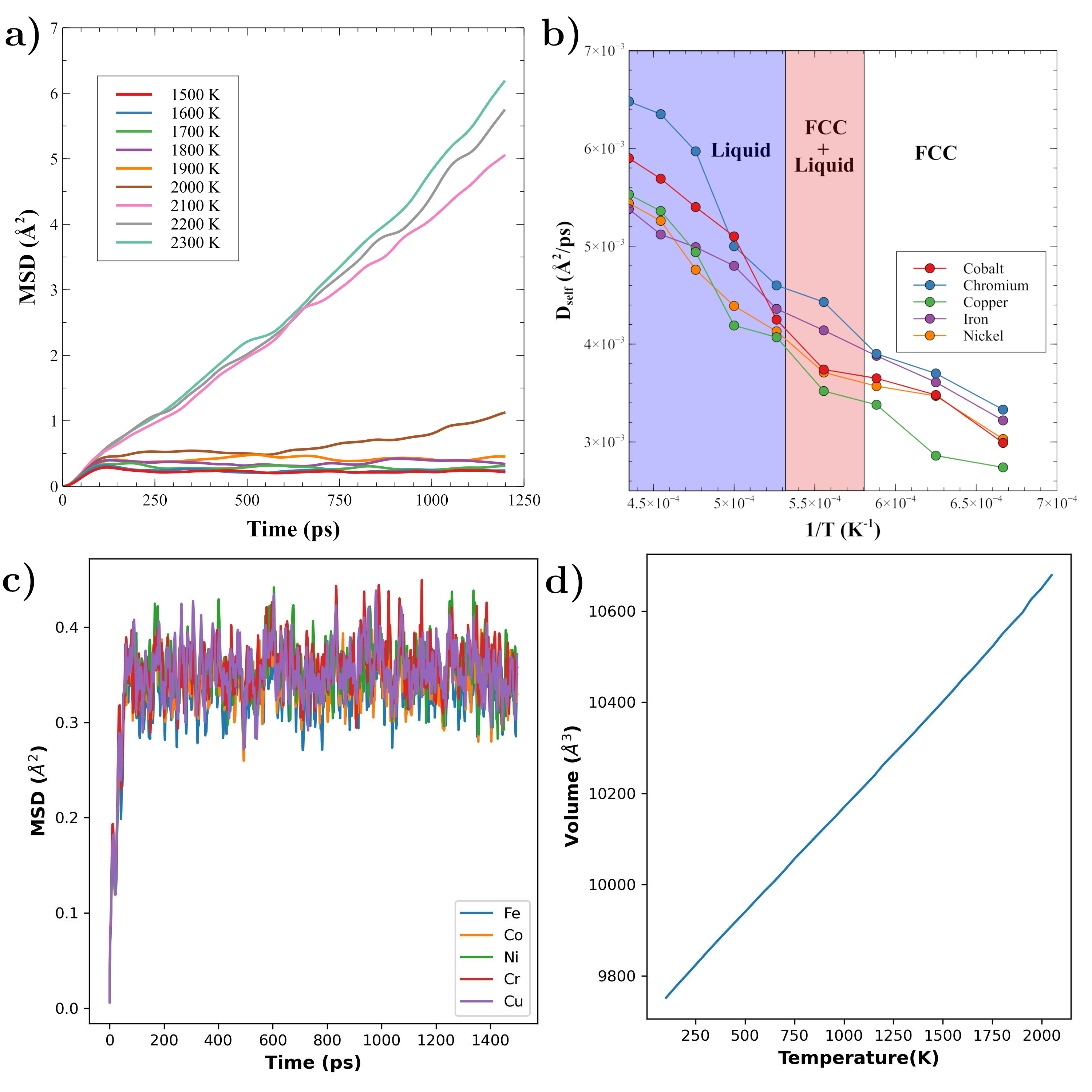}

\caption{\label{f:fig_diff} (a) Mean Square Displacement (MSD) as a function of simulation time for Cobalt in FNCCC using O-RFF potential. (b) Self-diffusion constants for each species in the alloy. (c) MSD curves for each 5 elements at 2300K as obtained by EAM potential. The horizontal plateau depicts a solid state at 2300K. (d) Temperature volume curve for FNCCC alloy as obtained using EAM potential.}
\end{figure}

Figure \ref{f:fig_diff}(a) shows the MSD of Cobalt in the FNCCC within the temperature range 1500K - 2300K. The MSD at lower temperatures have a linear region followed by a steady plateau. This is typical behaviour of solid state materials \cite{chen2013melting}. As the temperature increases, this plateau gradually disappears ( \(\sim\)1900K-2000K). MSD becomes a straight line with a single slope from 2100K onward indicating the complete transformation of the FNCCC to liquid state. This is in agreement with the evolution of crystalline ordering observed during the melting process as discussed in \ref{sec:melting}. MSD for other component species follows the same trend as this.


The coefficient of self-diffusion was calculated for each species using the O-RFF potentials. Self-diffusion constants ($D_{self}$) can be calculated from the asymptotic behaviour of the mean-square displacement (MSD) of atoms as follows
\begin{equation}
    \label{e:diff_eq}
    D_{self} = \frac{1}{6} \lim_{t\to\infty} \frac{d (MSD)}{dt}.
\end{equation}

Calculated diffusion constant values from the MSD are plotted in Figure \ref{f:fig_diff}(b). The absolute values of self-diffusion constants are of the same order of magnitude as reported in the literature \cite{neumann2011selfdiffusion}. This suggests that the O-RFF potential can be reliably used to perform calculations at elevated temperatures where the diffusive mechanisms are dominant.

A similar melting simulation was also performed using EAM potential [22] and corresponding results are shown in Figure \ref{f:fig_diff} (c-d). Figure \ref{f:fig_diff} (c) shows the variation of MSD for each element at 2400K. The horizontal plateau for each element indicates that the system is in solid state. Temperature-volume curve as obtained using EAM is a straight line with no visible transformation to liquid state Figure \ref{f:fig_diff} (d). This exhibits that EAM potential is not suitable for the simulations of FNCCC at elevated temperatures.

\subsection{Sparse potentials} \label{SparseResults}
The previous section details MD simulation results for the O-RFF potential and shows that property prediction through O-RFF agrees well with DFT results. A significant limitation with O-RFF is the number of parameters (15000) which significantly affects its scalability due to the run-time complexity of the MD simulations. To overcome this limitation, we trained a sparse version of the O-RFF potential using the RVM method, as detailed in section \ref{RVMSec}. Parameter estimation was performed using a greedy sequential optimization algorithm \cite{gd39}. Using the sparse model and sequential optimization resulted in 906 nonzero parameters, a reduction of 94\% from the original parameter space. It was noted that certain elemental interactions such as `Ni-Ni' were not selected by the algorithm. The sparse potential will be referred to as SparseRFF in the rest of the paper.

\begin{figure}[!ht]
\centering
\includegraphics[trim=-1 0.1 0.1 0.1,clip,scale=0.4]{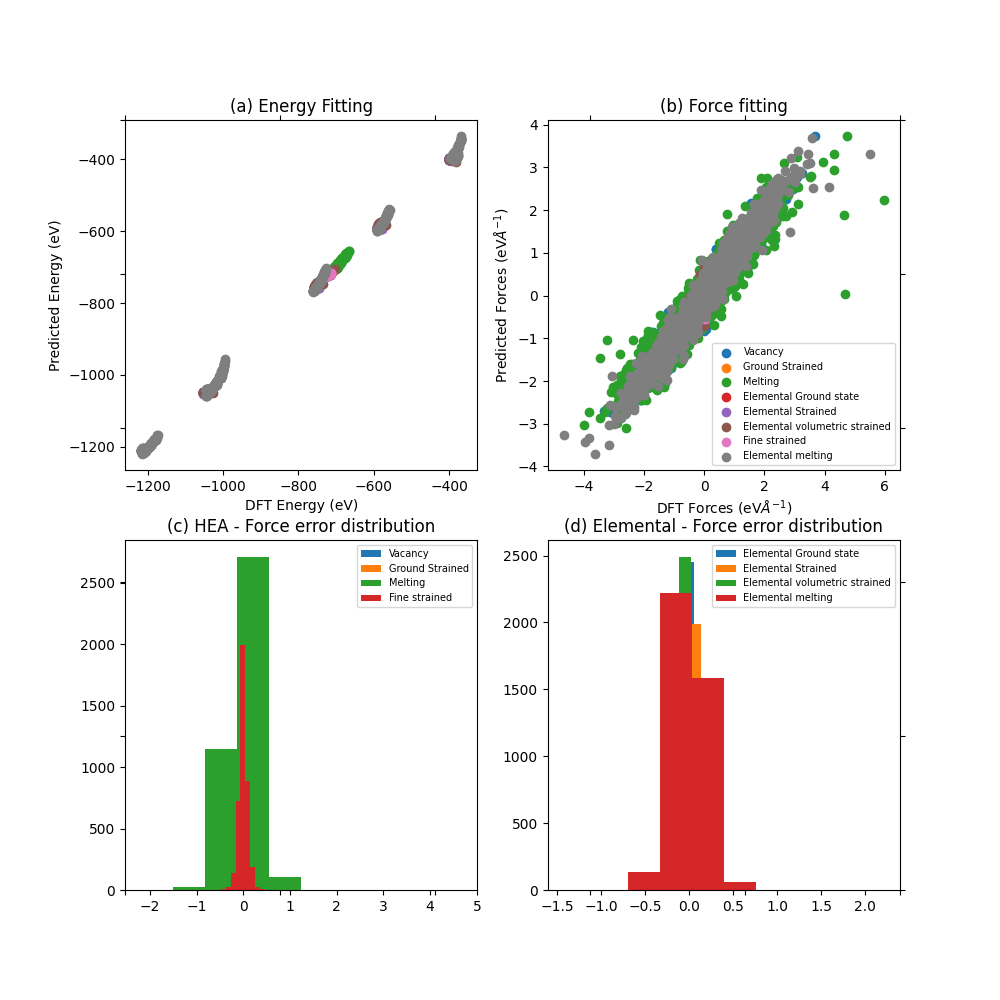}
\caption{\label{f:fig5} Comparison of DFT calculated energy force values to SparseRFF predicted energy (a), forces (b). Also shown are the corresponding force error distribution for FNCCC in (c) and elemental interactions in (d).}
\end{figure}

Figure \ref{f:fig5} shows the SparseRFF predicted energy and forces across all the datasets used in training. The overall error in energy and force prediction is higher than O-RFF but still, the majority of configurations have errors lying close to zero as compared to DFT-calculated energy/force values. Referring to Figure \ref{f:fig5}, force errors are higher in FNCCC (c) as compared to the elemental interactions (d). Force prediction for the data corresponding to elemental and FNCCC melting trajectories, showed a major discrepancy from DFT values. Apart from melting data, the force errors are less than 0.1$eV/\AA$ across all datasets, exhibiting good agreement with DFT values.

\subsubsection{MD Simulations}
Computed values of elastic constants by using SparseRFF potential are shown in Table \ref{Table:1}. Values of $C_{11}$ for all elements except Ni lie within 4\% of the DFT values. For Ni, there is a significant variation of 15\% in $C_{11}$ prediction. For all the elements, values for $C_{12}$ are within 9\% of the corresponding DFT results. The highest error in $C_{44}$ prediction is for Ni with variation as high as 16 GPa. $C_{44}$ predicted values for the rest of the elements is within 10 GPa or 11\% of the DFT values.

For each element, the bulk modulus, Young’s modulus and Poisson ratio are calculated using the equations described in Section \ref{mechPropSec}. The bulk modulus for all the elements except Ni is within 6\% of the DFT predicted value. For Co, bulk modulus as predicted by SparseRFF lied closer to DFT value as compared to the one computed using O-RFF. Young's modulus is also within 11\% of DFT calculated values. The discrepancy in bulk modulus and Young's modulus of Ni can be attributed to the over-prediction of the corresponding $C_{11}$ value. Finally, Poisson's ratio for all the elements is within 9\%.

Table \ref{Table:1} also shows the mechanical properties of the FNCCC alloy as computed using SparseRFF. All the properties except $C_{11}$ and $E$ are within 9\% of the corresponding DFT results. We observed that SparseRFF under-predicted $C_{11}$ by 20 GPa. This under-prediction led to the variance in Young's modulus prediction as well.

\section{Discussion}
This paper presented a computationally efficient IP that can model complex interactions of the FNCCC alloy as well its elemental constituents. The prediction of energy/forces is close to DFT accuracy leading to accurate property prediction through MD simulations. O-RFF can accurately model elastic constants, defect formation energies, melting behaviour and diffusion coefficients.

We also presented an approach for reducing the number of terms in the potential using Sparse Bayesian learning. The sparse potential was able to reproduce the basic properties while having a 94\% reduction in the number of parameters in the model. This reduction has a direct impact on the computational speed as shown by counting the number of FLOPS for local atomic energy calculations (ref. Table \ref{Table:3}). SparseRFF directly led to a reduction of 94\% in the number of FLOPS required. This is significant when the ML-based IP is used to simulate large-scale systems consisting of millions of atoms. 

\begin{table}[!ht]
\begin{tabular}{c c c}
Operation & O-RFF(15000) & SparseRFF(906) \\
\hline
$\omega^Tq_i^a$ &	15000/2 &	906/2 \\
$\cos(\omega^Tq_i^a)$ &	15000/2 &	906/2\\
$\sin(\omega^Tq_i^a)$ &	15000/2	 & 906/2\\
Energy computation  &	15000 &	906\\
Total number of operations & 37,500 &	2,265\\

\end{tabular}
\caption{\label{Table:3} The number of FLOPS required to compute energy using O-RFF and SparseRFF respectively. For O-RFF, the number of parameters is 15000 and for SparseRFF, the number of parameters is 906.}
\end{table}

It was also observed that certain interactions such as Ni-Ni were given zero weightage by the SparseRFF model. This means that these interactions can be completely removed during the prediction stage. This is in stark contrast to the spline-based potentials. The spline-based potentials need to maintain separate spline interpolation for each two-body and three-body interaction. It should be noted that this observation is data and property-dependent with no physical interpretation.

A key limitation with both O-RFF and SparseRFF is hyperparameter tuning. We performed hyperparameter tuning by validation through MD properties. By using proper calibration methods such as tools based on sequential optimization \cite{gd40}, the accuracy of these potentials can be improved. We also used uniform weight for each data point. The accuracy of certain properties can be improved by considering separate weights for the corresponding datasets. SparseRFF does not use any prior assumption about the physical system. If necessary, certain interactions can be given higher weight during basis selection. The major computational burden with both O-RFF and SparseRFF lies with the training data generation. In this regard, both potentials can benefit from active learning methods. Active learning can also help in the identification of the most important data points leading to a reduction in training set size. Finally, it has been shown that MD simulations are sensitive to IP parameter changes. To capture the uncertainty due to parameters in O-RFF, Bayesian methods as presented in \cite{gd41,gd42} can be used.

\section{Conclusion}
This paper explored the use of orthogonal random features for IP development of complex alloys consisting of five principal elements. Given the number of two-body and three-body interactions, it will be computationally inefficient and memory-demanding to use a kernel-based model such as GAP \cite{gd21}. The potential presented is based on generalized linear models and is more computationally efficient than standard kernel-based IPs. The potential was shown to predict mechanical as well as thermal properties with DFT accuracy. During the melting, the disordering of the atoms was investigated. Self-diffusion in the alloy was modelled at temperatures close to the melting point. We also presented a novel sparse Bayesian learning-based potential that reduced the computational effort by 94\%. To the best of our knowledge, sparse Bayesian learning-based potentials have not been explored before in the literature. We showed that with the reduced parameter space, SparseRFF can predict the final properties within 10\% of the DFT values.

\section*{Acknowledgements}
This research is supported by grants from Natural Sciences and Engineering Research Council of Canada, Hart Professorship, Canada Research Chairs program, the University of Toronto and Digital Research Alliance of Canada.

\bibliography{abu_biblio}


\begin{thebibliography}{61}
\ifx \bisbn   \undefined \def \bisbn  #1{ISBN #1}\fi
\ifx \binits  \undefined \def \binits#1{#1}\fi
\ifx \bauthor  \undefined \def \bauthor#1{#1}\fi
\ifx \batitle  \undefined \def \batitle#1{#1}\fi
\ifx \bjtitle  \undefined \def \bjtitle#1{#1}\fi
\ifx \bvolume  \undefined \def \bvolume#1{\textbf{#1}}\fi
\ifx \byear  \undefined \def \byear#1{#1}\fi
\ifx \bissue  \undefined \def \bissue#1{#1}\fi
\ifx \bfpage  \undefined \def \bfpage#1{#1}\fi
\ifx \blpage  \undefined \def \blpage #1{#1}\fi
\ifx \burl  \undefined \def \burl#1{\textsf{#1}}\fi
\ifx \doiurl  \undefined \def \doiurl#1{\url{https://doi.org/#1}}\fi
\ifx \betal  \undefined \def \betal{\textit{et al.}}\fi
\ifx \binstitute  \undefined \def \binstitute#1{#1}\fi
\ifx \binstitutionaled  \undefined \def \binstitutionaled#1{#1}\fi
\ifx \bctitle  \undefined \def \bctitle#1{#1}\fi
\ifx \beditor  \undefined \def \beditor#1{#1}\fi
\ifx \bpublisher  \undefined \def \bpublisher#1{#1}\fi
\ifx \bbtitle  \undefined \def \bbtitle#1{#1}\fi
\ifx \bedition  \undefined \def \bedition#1{#1}\fi
\ifx \bseriesno  \undefined \def \bseriesno#1{#1}\fi
\ifx \blocation  \undefined \def \blocation#1{#1}\fi
\ifx \bsertitle  \undefined \def \bsertitle#1{#1}\fi
\ifx \bsnm \undefined \def \bsnm#1{#1}\fi
\ifx \bsuffix \undefined \def \bsuffix#1{#1}\fi
\ifx \bparticle \undefined \def \bparticle#1{#1}\fi
\ifx \barticle \undefined \def \barticle#1{#1}\fi
\bibcommenthead
\ifx \bconfdate \undefined \def \bconfdate #1{#1}\fi
\ifx \botherref \undefined \def \botherref #1{#1}\fi
\ifx \url \undefined \def \url#1{\textsf{#1}}\fi
\ifx \bchapter \undefined \def \bchapter#1{#1}\fi
\ifx \bbook \undefined \def \bbook#1{#1}\fi
\ifx \bcomment \undefined \def \bcomment#1{#1}\fi
\ifx \oauthor \undefined \def \oauthor#1{#1}\fi
\ifx \citeauthoryear \undefined \def \citeauthoryear#1{#1}\fi
\ifx \endbibitem  \undefined \def \endbibitem {}\fi
\ifx \bconflocation  \undefined \def \bconflocation#1{#1}\fi
\ifx \arxivurl  \undefined \def \arxivurl#1{\textsf{#1}}\fi
\csname PreBibitemsHook\endcsname

\bibitem{gd1}
\begin{barticle}
\bauthor{\bsnm{George}, \binits{E.P.}},
\bauthor{\bsnm{Raabe}, \binits{D.}},
\bauthor{\bsnm{Ritchie}, \binits{R.O.}}:
\batitle{High-entropy alloys}.
\bjtitle{Nature reviews materials}
\bvolume{4}(\bissue{8}),
\bfpage{515}--\blpage{534}
(\byear{2019})
\end{barticle}
\endbibitem

\bibitem{gd4}
\begin{barticle}
\bauthor{\bsnm{George}, \binits{E.P.}},
\bauthor{\bsnm{Curtin}, \binits{W.}},
\bauthor{\bsnm{Tasan}, \binits{C.C.}}:
\batitle{High entropy alloys: A focused review of mechanical properties and
  deformation mechanisms}.
\bjtitle{Acta Materialia}
\bvolume{188},
\bfpage{435}--\blpage{474}
(\byear{2020})
\end{barticle}
\endbibitem

\bibitem{gd5}
\begin{barticle}
\bauthor{\bsnm{Chen}, \binits{J.}},
\bauthor{\bsnm{Zhou}, \binits{X.}},
\bauthor{\bsnm{Wang}, \binits{W.}},
\bauthor{\bsnm{Liu}, \binits{B.}},
\bauthor{\bsnm{Lv}, \binits{Y.}},
\bauthor{\bsnm{Yang}, \binits{W.}},
\bauthor{\bsnm{Xu}, \binits{D.}},
\bauthor{\bsnm{Liu}, \binits{Y.}}:
\batitle{A review on fundamental of high entropy alloys with promising
  high--temperature properties}.
\bjtitle{Journal of Alloys and Compounds}
\bvolume{760},
\bfpage{15}--\blpage{30}
(\byear{2018})
\end{barticle}
\endbibitem

\bibitem{anand2023recent}
\begin{botherref}
\oauthor{\bsnm{Anand}, \binits{A.}},
\oauthor{\bsnm{Liu}, \binits{S.-J.}},
\oauthor{\bsnm{Singh}, \binits{C.V.}}:
Recent advances in computational design of structural multi-principal element
  alloys.
iScience,
107751
(2023)
\end{botherref}
\endbibitem

\bibitem{gd2}
\begin{barticle}
\bauthor{\bsnm{Toda-Caraballo}, \binits{I.}},
\bauthor{\bsnm{Wr{\'o}bel}, \binits{J.S.}},
\bauthor{\bsnm{Nguyen-Manh}, \binits{D.}},
\bauthor{\bsnm{P{\'e}rez}, \binits{P.}},
\bauthor{\bsnm{Rivera-D{\'\i}az-del-Castillo}, \binits{P.}}:
\batitle{Simulation and modeling in high entropy alloys}.
\bjtitle{JOM}
\bvolume{69}(\bissue{11}),
\bfpage{2137}--\blpage{2149}
(\byear{2017})
\end{barticle}
\endbibitem

\bibitem{gd3}
\begin{barticle}
\bauthor{\bsnm{Ferrari}, \binits{A.}},
\bauthor{\bsnm{Dutta}, \binits{B.}},
\bauthor{\bsnm{Gubaev}, \binits{K.}},
\bauthor{\bsnm{Ikeda}, \binits{Y.}},
\bauthor{\bsnm{Srinivasan}, \binits{P.}},
\bauthor{\bsnm{Grabowski}, \binits{B.}},
\bauthor{\bsnm{K{\"o}rmann}, \binits{F.}}:
\batitle{Frontiers in atomistic simulations of high entropy alloys}.
\bjtitle{Journal of Applied Physics}
\bvolume{128}(\bissue{15}),
\bfpage{150901}
(\byear{2020})
\end{barticle}
\endbibitem

\bibitem{gd6}
\begin{bbook}
\bauthor{\bsnm{LeSar}, \binits{R.}}:
\bbtitle{Introduction to Computational Materials Science: Fundamentals to
  Applications}.
\bpublisher{Cambridge University Press}, \blocation{???}
(\byear{2013})
\end{bbook}
\endbibitem

\bibitem{gd7}
\begin{barticle}
\bauthor{\bsnm{Antillon}, \binits{E.}},
\bauthor{\bsnm{Woodward}, \binits{C.}},
\bauthor{\bsnm{Rao}, \binits{S.}},
\bauthor{\bsnm{Akdim}, \binits{B.}},
\bauthor{\bsnm{Parthasarathy}, \binits{T.}}:
\batitle{Chemical short range order strengthening in a model fcc high entropy
  alloy}.
\bjtitle{Acta Materialia}
\bvolume{190},
\bfpage{29}--\blpage{42}
(\byear{2020})
\end{barticle}
\endbibitem

\bibitem{gd8}
\begin{barticle}
\bauthor{\bsnm{Farkas}, \binits{D.}},
\bauthor{\bsnm{Caro}, \binits{A.}}:
\batitle{Model interatomic potentials and lattice strain in a high-entropy
  alloy}.
\bjtitle{Journal of Materials Research}
\bvolume{33}(\bissue{19}),
\bfpage{3218}--\blpage{3225}
(\byear{2018})
\end{barticle}
\endbibitem

\bibitem{gd9}
\begin{barticle}
\bauthor{\bsnm{Farkas}, \binits{D.}},
\bauthor{\bsnm{Caro}, \binits{A.}}:
\batitle{Model interatomic potentials for fe--ni--cr--co--al high-entropy
  alloys}.
\bjtitle{Journal of Materials Research}
\bvolume{35}(\bissue{22}),
\bfpage{3031}--\blpage{3040}
(\byear{2020})
\end{barticle}
\endbibitem

\bibitem{gd10}
\begin{barticle}
\bauthor{\bsnm{Deluigi}, \binits{O.}},
\bauthor{\bsnm{Pasianot}, \binits{R.}},
\bauthor{\bsnm{Valencia}, \binits{F.}},
\bauthor{\bsnm{Caro}, \binits{A.}},
\bauthor{\bsnm{Farkas}, \binits{D.}},
\bauthor{\bsnm{Bringa}, \binits{E.}}:
\batitle{Simulations of primary damage in a high entropy alloy: Probing
  enhanced radiation resistance}.
\bjtitle{Acta Materialia}
\bvolume{213},
\bfpage{116951}
(\byear{2021})
\end{barticle}
\endbibitem

\bibitem{gd11}
\begin{barticle}
\bauthor{\bsnm{Huang}, \binits{X.}},
\bauthor{\bsnm{Liu}, \binits{L.}},
\bauthor{\bsnm{Duan}, \binits{X.}},
\bauthor{\bsnm{Liao}, \binits{W.}},
\bauthor{\bsnm{Huang}, \binits{J.}},
\bauthor{\bsnm{Sun}, \binits{H.}},
\bauthor{\bsnm{Yu}, \binits{C.}}:
\batitle{Atomistic simulation of chemical short-range order in hfnbtazr high
  entropy alloy based on a newly-developed interatomic potential}.
\bjtitle{Materials \& Design}
\bvolume{202},
\bfpage{109560}
(\byear{2021})
\end{barticle}
\endbibitem

\bibitem{gd12}
\begin{barticle}
\bauthor{\bsnm{Liu}, \binits{J.}}:
\batitle{Molecular dynamic study of temperature dependence of mechanical
  properties and plastic inception of cocrcufeni high-entropy alloy}.
\bjtitle{Physics Letters A}
\bvolume{384}(\bissue{22}),
\bfpage{126516}
(\byear{2020})
\end{barticle}
\endbibitem

\bibitem{gd13}
\begin{barticle}
\bauthor{\bsnm{Utt}, \binits{D.}},
\bauthor{\bsnm{Stukowski}, \binits{A.}},
\bauthor{\bsnm{Albe}, \binits{K.}}:
\batitle{Grain boundary structure and mobility in high-entropy alloys: A
  comparative molecular dynamics study on a $\sigma$11 symmetrical tilt grain
  boundary in face-centered cubic cunicofe}.
\bjtitle{Acta Materialia}
\bvolume{186},
\bfpage{11}--\blpage{19}
(\byear{2020})
\end{barticle}
\endbibitem

\bibitem{gd14}
\begin{barticle}
\bauthor{\bsnm{Choi}, \binits{W.-M.}},
\bauthor{\bsnm{Jo}, \binits{Y.H.}},
\bauthor{\bsnm{Sohn}, \binits{S.S.}},
\bauthor{\bsnm{Lee}, \binits{S.}},
\bauthor{\bsnm{Lee}, \binits{B.-J.}}:
\batitle{Understanding the physical metallurgy of the cocrfemnni high-entropy
  alloy: an atomistic simulation study}.
\bjtitle{npj Computational Materials}
\bvolume{4}(\bissue{1}),
\bfpage{1}--\blpage{9}
(\byear{2018})
\end{barticle}
\endbibitem

\bibitem{gd15}
\begin{barticle}
\bauthor{\bsnm{Antillon}, \binits{E.}},
\bauthor{\bsnm{Woodward}, \binits{C.}},
\bauthor{\bsnm{Rao}, \binits{S.}},
\bauthor{\bsnm{Akdim}, \binits{B.}},
\bauthor{\bsnm{Parthasarathy}, \binits{T.}}:
\batitle{Chemical short range order strengthening in a model fcc high entropy
  alloy}.
\bjtitle{Acta Materialia}
\bvolume{190},
\bfpage{29}--\blpage{42}
(\byear{2020})
\end{barticle}
\endbibitem

\bibitem{gd16}
\begin{barticle}
\bauthor{\bsnm{Daramola}, \binits{A.}},
\bauthor{\bsnm{Bonny}, \binits{G.}},
\bauthor{\bsnm{Adjanor}, \binits{G.}},
\bauthor{\bsnm{Domain}, \binits{C.}},
\bauthor{\bsnm{Monnet}, \binits{G.}},
\bauthor{\bsnm{Fraczkiewicz}, \binits{A.}}:
\batitle{Development of a plasticity-oriented interatomic potential for
  crfemnni high entropy alloys}.
\bjtitle{Computational Materials Science}
\bvolume{203},
\bfpage{111165}
(\byear{2022})
\end{barticle}
\endbibitem

\bibitem{gd17}
\begin{barticle}
\bauthor{\bsnm{Mueller}, \binits{T.}},
\bauthor{\bsnm{Hernandez}, \binits{A.}},
\bauthor{\bsnm{Wang}, \binits{C.}}:
\batitle{Machine learning for interatomic potential models}.
\bjtitle{The Journal of chemical physics}
\bvolume{152}(\bissue{5}),
\bfpage{050902}
(\byear{2020})
\end{barticle}
\endbibitem

\bibitem{gd18}
\begin{barticle}
\bauthor{\bsnm{Deringer}, \binits{V.L.}},
\bauthor{\bsnm{Caro}, \binits{M.A.}},
\bauthor{\bsnm{Cs{\'a}nyi}, \binits{G.}}:
\batitle{Machine learning interatomic potentials as emerging tools for
  materials science}.
\bjtitle{Advanced Materials}
\bvolume{31}(\bissue{46}),
\bfpage{1902765}
(\byear{2019})
\end{barticle}
\endbibitem

\bibitem{gd20}
\begin{barticle}
\bauthor{\bsnm{Behler}, \binits{J.}}:
\batitle{Four generations of high-dimensional neural network potentials}.
\bjtitle{Chemical Reviews}
\bvolume{121}(\bissue{16}),
\bfpage{10037}--\blpage{10072}
(\byear{2021})
\end{barticle}
\endbibitem

\bibitem{gd21}
\begin{barticle}
\bauthor{\bsnm{Deringer}, \binits{V.L.}},
\bauthor{\bsnm{Bart{\'o}k}, \binits{A.P.}},
\bauthor{\bsnm{Bernstein}, \binits{N.}},
\bauthor{\bsnm{Wilkins}, \binits{D.M.}},
\bauthor{\bsnm{Ceriotti}, \binits{M.}},
\bauthor{\bsnm{Cs{\'a}nyi}, \binits{G.}}:
\batitle{Gaussian process regression for materials and molecules}.
\bjtitle{Chemical Reviews}
\bvolume{121}(\bissue{16}),
\bfpage{10073}--\blpage{10141}
(\byear{2021})
\end{barticle}
\endbibitem

\bibitem{gd23}
\begin{barticle}
\bauthor{\bsnm{Thompson}, \binits{A.P.}},
\bauthor{\bsnm{Swiler}, \binits{L.P.}},
\bauthor{\bsnm{Trott}, \binits{C.R.}},
\bauthor{\bsnm{Foiles}, \binits{S.M.}},
\bauthor{\bsnm{Tucker}, \binits{G.J.}}:
\batitle{Spectral neighbor analysis method for automated generation of
  quantum-accurate interatomic potentials}.
\bjtitle{Journal of Computational Physics}
\bvolume{285},
\bfpage{316}--\blpage{330}
(\byear{2015})
\end{barticle}
\endbibitem

\bibitem{gd24}
\begin{barticle}
\bauthor{\bsnm{Goryaeva}, \binits{A.M.}},
\bauthor{\bsnm{Maillet}, \binits{J.-B.}},
\bauthor{\bsnm{Marinica}, \binits{M.-C.}}:
\batitle{Towards better efficiency of interatomic linear machine learning
  potentials}.
\bjtitle{Computational Materials Science}
\bvolume{166},
\bfpage{200}--\blpage{209}
(\byear{2019})
\end{barticle}
\endbibitem

\bibitem{gd29}
\begin{barticle}
\bauthor{\bsnm{Thompson}, \binits{A.P.}},
\bauthor{\bsnm{Swiler}, \binits{L.P.}},
\bauthor{\bsnm{Trott}, \binits{C.R.}},
\bauthor{\bsnm{Foiles}, \binits{S.M.}},
\bauthor{\bsnm{Tucker}, \binits{G.J.}}:
\batitle{Spectral neighbor analysis method for automated generation of
  quantum-accurate interatomic potentials}.
\bjtitle{Journal of Computational Physics}
\bvolume{285},
\bfpage{316}--\blpage{330}
(\byear{2015})
\end{barticle}
\endbibitem

\bibitem{gd19}
\begin{barticle}
\bauthor{\bsnm{Mishin}, \binits{Y.}}:
\batitle{Machine-learning interatomic potentials for materials science}.
\bjtitle{Acta Materialia}
\bvolume{214},
\bfpage{116980}
(\byear{2021})
\end{barticle}
\endbibitem

\bibitem{gd22}
\begin{barticle}
\bauthor{\bsnm{Zuo}, \binits{Y.}},
\bauthor{\bsnm{Chen}, \binits{C.}},
\bauthor{\bsnm{Li}, \binits{X.}},
\bauthor{\bsnm{Deng}, \binits{Z.}},
\bauthor{\bsnm{Chen}, \binits{Y.}},
\bauthor{\bsnm{Behler}, \binits{J.}},
\bauthor{\bsnm{Cs{\'a}nyi}, \binits{G.}},
\bauthor{\bsnm{Shapeev}, \binits{A.V.}},
\bauthor{\bsnm{Thompson}, \binits{A.P.}},
\bauthor{\bsnm{Wood}, \binits{M.A.}}, \betal:
\batitle{Performance and cost assessment of machine learning interatomic
  potentials}.
\bjtitle{The Journal of Physical Chemistry A}
\bvolume{124}(\bissue{4}),
\bfpage{731}--\blpage{745}
(\byear{2020})
\end{barticle}
\endbibitem

\bibitem{gd25}
\begin{barticle}
\bauthor{\bsnm{Vandermause}, \binits{J.}},
\bauthor{\bsnm{Torrisi}, \binits{S.B.}},
\bauthor{\bsnm{Batzner}, \binits{S.}},
\bauthor{\bsnm{Xie}, \binits{Y.}},
\bauthor{\bsnm{Sun}, \binits{L.}},
\bauthor{\bsnm{Kolpak}, \binits{A.M.}},
\bauthor{\bsnm{Kozinsky}, \binits{B.}}:
\batitle{On-the-fly active learning of interpretable bayesian force fields for
  atomistic rare events}.
\bjtitle{npj Computational Materials}
\bvolume{6}(\bissue{1}),
\bfpage{1}--\blpage{11}
(\byear{2020})
\end{barticle}
\endbibitem

\bibitem{gd26}
\begin{barticle}
\bauthor{\bsnm{Byggm{\"a}star}, \binits{J.}},
\bauthor{\bsnm{Nordlund}, \binits{K.}},
\bauthor{\bsnm{Djurabekova}, \binits{F.}}:
\batitle{Modeling refractory high-entropy alloys with efficient machine-learned
  interatomic potentials: Defects and segregation}.
\bjtitle{Physical Review B}
\bvolume{104}(\bissue{10}),
\bfpage{104101}
(\byear{2021})
\end{barticle}
\endbibitem

\bibitem{gd27}
\begin{botherref}
\oauthor{\bsnm{Byggm{\"a}star}, \binits{J.}},
\oauthor{\bsnm{Nordlund}, \binits{K.}},
\oauthor{\bsnm{Djurabekova}, \binits{F.}}:
Simple machine-learned interatomic potentials for complex alloys.
arXiv preprint arXiv:2203.08458
(2022)
\end{botherref}
\endbibitem

\bibitem{gd28}
\begin{barticle}
\bauthor{\bsnm{Li}, \binits{X.-G.}},
\bauthor{\bsnm{Chen}, \binits{C.}},
\bauthor{\bsnm{Zheng}, \binits{H.}},
\bauthor{\bsnm{Zuo}, \binits{Y.}},
\bauthor{\bsnm{Ong}, \binits{S.P.}}:
\batitle{Complex strengthening mechanisms in the nbmotaw multi-principal
  element alloy}.
\bjtitle{npj Computational Materials}
\bvolume{6}(\bissue{1}),
\bfpage{1}--\blpage{10}
(\byear{2020})
\end{barticle}
\endbibitem

\bibitem{gd30}
\begin{barticle}
\bauthor{\bsnm{Shapeev}, \binits{A.}}:
\batitle{Accurate representation of formation energies of crystalline alloys
  with many components}.
\bjtitle{Computational Materials Science}
\bvolume{139},
\bfpage{26}--\blpage{30}
(\byear{2017})
\end{barticle}
\endbibitem

\bibitem{gd31}
\begin{barticle}
\bauthor{\bsnm{Meshkov}, \binits{E.}},
\bauthor{\bsnm{Novoselov}, \binits{I.}},
\bauthor{\bsnm{Shapeev}, \binits{A.}},
\bauthor{\bsnm{Yanilkin}, \binits{A.}}:
\batitle{Sublattice formation in cocrfeni high-entropy alloy}.
\bjtitle{Intermetallics}
\bvolume{112},
\bfpage{106542}
(\byear{2019})
\end{barticle}
\endbibitem

\bibitem{gd32}
\begin{barticle}
\bauthor{\bsnm{Kostiuchenko}, \binits{T.}},
\bauthor{\bsnm{K{\"o}rmann}, \binits{F.}},
\bauthor{\bsnm{Neugebauer}, \binits{J.}},
\bauthor{\bsnm{Shapeev}, \binits{A.}}:
\batitle{Impact of lattice relaxations on phase transitions in a high-entropy
  alloy studied by machine-learning potentials}.
\bjtitle{npj Computational Materials}
\bvolume{5}(\bissue{1}),
\bfpage{1}--\blpage{7}
(\byear{2019})
\end{barticle}
\endbibitem

\bibitem{gd33}
\begin{barticle}
\bauthor{\bsnm{Shapeev}, \binits{A.V.}}:
\batitle{Moment tensor potentials: A class of systematically improvable
  interatomic potentials}.
\bjtitle{Multiscale Modeling \& Simulation}
\bvolume{14}(\bissue{3}),
\bfpage{1153}--\blpage{1173}
(\byear{2016})
\end{barticle}
\endbibitem

\bibitem{gd34}
\begin{barticle}
\bauthor{\bsnm{Jafary-Zadeh}, \binits{M.}},
\bauthor{\bsnm{Khoo}, \binits{K.H.}},
\bauthor{\bsnm{Laskowski}, \binits{R.}},
\bauthor{\bsnm{Branicio}, \binits{P.S.}},
\bauthor{\bsnm{Shapeev}, \binits{A.V.}}:
\batitle{Applying a machine learning interatomic potential to unravel the
  effects of local lattice distortion on the elastic properties of
  multi-principal element alloys}.
\bjtitle{Journal of Alloys and Compounds}
\bvolume{803},
\bfpage{1054}--\blpage{1062}
(\byear{2019})
\end{barticle}
\endbibitem

\bibitem{gd35}
\begin{barticle}
\bauthor{\bsnm{Grabowski}, \binits{B.}},
\bauthor{\bsnm{Ikeda}, \binits{Y.}},
\bauthor{\bsnm{Srinivasan}, \binits{P.}},
\bauthor{\bsnm{K{\"o}rmann}, \binits{F.}},
\bauthor{\bsnm{Freysoldt}, \binits{C.}},
\bauthor{\bsnm{Duff}, \binits{A.I.}},
\bauthor{\bsnm{Shapeev}, \binits{A.}},
\bauthor{\bsnm{Neugebauer}, \binits{J.}}:
\batitle{Ab initio vibrational free energies including anharmonicity for
  multicomponent alloys}.
\bjtitle{npj Computational Materials}
\bvolume{5}(\bissue{1}),
\bfpage{1}--\blpage{6}
(\byear{2019})
\end{barticle}
\endbibitem

\bibitem{gdNPJRFF}
\begin{barticle}
\bauthor{\bsnm{Dhaliwal}, \binits{G.}},
\bauthor{\bsnm{Nair}, \binits{P.B.}},
\bauthor{\bsnm{Singh}, \binits{C.V.}}:
\batitle{Machine learned interatomic potentials using random features}.
\bjtitle{npj Computational Materials}
\bvolume{8}(\bissue{1}),
\bfpage{1}--\blpage{10}
(\byear{2022})
\end{barticle}
\endbibitem

\bibitem{gd36}
\begin{botherref}
\oauthor{\bsnm{Rahimi}, \binits{A.}},
\oauthor{\bsnm{Recht}, \binits{B.}}:
Random features for large-scale kernel machines.
Advances in neural information processing systems
\textbf{20}
(2007)
\end{botherref}
\endbibitem

\bibitem{gd37}
\begin{barticle}
\bauthor{\bsnm{Tipping}, \binits{M.E.}}:
\batitle{Sparse bayesian learning and the relevance vector machine}.
\bjtitle{Journal of machine learning research}
\bvolume{1}(\bissue{Jun}),
\bfpage{211}--\blpage{244}
(\byear{2001})
\end{barticle}
\endbibitem

\bibitem{gd38}
\begin{bchapter}
\bauthor{\bsnm{Tipping}, \binits{M.E.}}:
\bctitle{Bayesian inference: An introduction to principles and practice in
  machine learning}.
In: \bbtitle{Summer School on Machine Learning},
pp. \bfpage{41}--\blpage{62}
(\byear{2003}).
\bcomment{Springer}
\end{bchapter}
\endbibitem

\bibitem{gd39}
\begin{bchapter}
\bauthor{\bsnm{Tipping}, \binits{M.E.}},
\bauthor{\bsnm{Faul}, \binits{A.C.}}:
\bctitle{Fast marginal likelihood maximisation for sparse bayesian models}.
In: \bbtitle{International Workshop on Artificial Intelligence and Statistics},
pp. \bfpage{276}--\blpage{283}
(\byear{2003}).
\bcomment{PMLR}
\end{bchapter}
\endbibitem

\bibitem{kresse1993ab}
\begin{barticle}
\bauthor{\bsnm{Kresse}, \binits{G.}},
\bauthor{\bsnm{Hafner}, \binits{J.}}:
\batitle{Ab initio molecular dynamics for liquid metals}.
\bjtitle{Physical review B}
\bvolume{47}(\bissue{1}),
\bfpage{558}
(\byear{1993})
\end{barticle}
\endbibitem

\bibitem{kresse1996phys}
\begin{barticle}
\bauthor{\bsnm{Kresse}, \binits{G.}},
\bauthor{\bsnm{Furthm{\"u}ller}, \binits{J.}}:
\batitle{Efficient iterative schemes for ab initio total-energy calculations
  using a plane-wave basis set}.
\bjtitle{Physical review B}
\bvolume{54}(\bissue{16}),
\bfpage{11169}
(\byear{1996})
\end{barticle}
\endbibitem

\bibitem{blochl1994projector}
\begin{barticle}
\bauthor{\bsnm{Bl{\"o}chl}, \binits{P.E.}}:
\batitle{Projector augmented-wave method}.
\bjtitle{Physical review B}
\bvolume{50}(\bissue{24}),
\bfpage{17953}
(\byear{1994})
\end{barticle}
\endbibitem

\bibitem{kresse1999ultrasoft}
\begin{barticle}
\bauthor{\bsnm{Kresse}, \binits{G.}},
\bauthor{\bsnm{Joubert}, \binits{D.}}:
\batitle{From ultrasoft pseudopotentials to the projector augmented-wave
  method}.
\bjtitle{Physical review b}
\bvolume{59}(\bissue{3}),
\bfpage{1758}
(\byear{1999})
\end{barticle}
\endbibitem

\bibitem{perdew1996generalized}
\begin{barticle}
\bauthor{\bsnm{Perdew}, \binits{J.P.}},
\bauthor{\bsnm{Burke}, \binits{K.}},
\bauthor{\bsnm{Ernzerhof}, \binits{M.}}:
\batitle{Generalized gradient approximation made simple}.
\bjtitle{Physical review letters}
\bvolume{77}(\bissue{18}),
\bfpage{3865}
(\byear{1996})
\end{barticle}
\endbibitem

\bibitem{monkhorst1976special}
\begin{barticle}
\bauthor{\bsnm{Monkhorst}, \binits{H.J.}},
\bauthor{\bsnm{Pack}, \binits{J.D.}}:
\batitle{Special points for brillouin-zone integrations}.
\bjtitle{Physical review B}
\bvolume{13}(\bissue{12}),
\bfpage{5188}
(\byear{1976})
\end{barticle}
\endbibitem

\bibitem{aseRef}
\begin{barticle}
\bauthor{\bsnm{Larsen}, \binits{A.H.}},
\bauthor{\bsnm{Mortensen}, \binits{J.J.}},
\bauthor{\bsnm{Blomqvist}, \binits{J.}},
\bauthor{\bsnm{Castelli}, \binits{I.E.}},
\bauthor{\bsnm{Christensen}, \binits{R.}},
\bauthor{\bsnm{Du{\l}ak}, \binits{M.}},
\bauthor{\bsnm{Friis}, \binits{J.}},
\bauthor{\bsnm{Groves}, \binits{M.N.}},
\bauthor{\bsnm{Hammer}, \binits{B.}},
\bauthor{\bsnm{Hargus}, \binits{C.}}, \betal:
\batitle{The atomic simulation environment—a python library for working with
  atoms}.
\bjtitle{Journal of Physics: Condensed Matter}
\bvolume{29}(\bissue{27}),
\bfpage{273002}
(\byear{2017})
\end{barticle}
\endbibitem

\bibitem{zhang2021statistical}
\begin{barticle}
\bauthor{\bsnm{Zhang}, \binits{Y.}},
\bauthor{\bsnm{Manzoor}, \binits{A.}},
\bauthor{\bsnm{Jiang}, \binits{C.}},
\bauthor{\bsnm{Aidhy}, \binits{D.}},
\bauthor{\bsnm{Schwen}, \binits{D.}}:
\batitle{A statistical approach for atomistic calculations of vacancy formation
  energy and chemical potentials in concentrated solid-solution alloys}.
\bjtitle{Computational Materials Science}
\bvolume{190},
\bfpage{110308}
(\byear{2021})
\end{barticle}
\endbibitem

\bibitem{ojha2014twinning}
\begin{barticle}
\bauthor{\bsnm{Ojha}, \binits{A.}},
\bauthor{\bsnm{Sehitoglu}, \binits{H.}}:
\batitle{Twinning stress prediction in bcc metals and alloys}.
\bjtitle{Philosophical magazine letters}
\bvolume{94}(\bissue{10}),
\bfpage{647}--\blpage{657}
(\byear{2014})
\end{barticle}
\endbibitem

\bibitem{EAM_IP}
\begin{barticle}
\bauthor{\bsnm{Deluigi}, \binits{O.}},
\bauthor{\bsnm{Pasianot}, \binits{R.}},
\bauthor{\bsnm{Valencia}, \binits{F.}},
\bauthor{\bsnm{Caro}, \binits{A.}},
\bauthor{\bsnm{Farkas}, \binits{D.}},
\bauthor{\bsnm{Bringa}, \binits{E.}}:
\batitle{Simulations of primary damage in a high entropy alloy: Probing
  enhanced radiation resistance}.
\bjtitle{Acta Materialia}
\bvolume{213},
\bfpage{116951}
(\byear{2021})
\end{barticle}
\endbibitem

\bibitem{hirel2015atomsk}
\begin{barticle}
\bauthor{\bsnm{Hirel}, \binits{P.}}:
\batitle{Atomsk: A tool for manipulating and converting atomic data files}.
\bjtitle{Computer Physics Communications}
\bvolume{197},
\bfpage{212}--\blpage{219}
(\byear{2015})
\end{barticle}
\endbibitem

\bibitem{ovitoRef}
\begin{barticle}
\bauthor{\bsnm{Stukowski}, \binits{A.}}:
\batitle{Visualization and analysis of atomistic simulation data with
  ovito--the open visualization tool}.
\bjtitle{Modelling and simulation in materials science and engineering}
\bvolume{18}(\bissue{1}),
\bfpage{015012}
(\byear{2009})
\end{barticle}
\endbibitem

\bibitem{meltFNCCC}
\begin{barticle}
\bauthor{\bsnm{Wang}, \binits{X.}},
\bauthor{\bsnm{Zhang}, \binits{Y.}},
\bauthor{\bsnm{Qiao}, \binits{Y.}},
\bauthor{\bsnm{Chen}, \binits{G.}}:
\batitle{Novel microstructure and properties of multicomponent cocrcufenitix
  alloys}.
\bjtitle{Intermetallics}
\bvolume{15}(\bissue{3}),
\bfpage{357}--\blpage{362}
(\byear{2007})
\end{barticle}
\endbibitem

\bibitem{mei1990molecular}
\begin{barticle}
\bauthor{\bsnm{Mei}, \binits{J.}},
\bauthor{\bsnm{Davenport}, \binits{J.}}:
\batitle{Molecular-dynamics study of self-diffusion in liquid transition
  metals}.
\bjtitle{Physical Review B}
\bvolume{42}(\bissue{15}),
\bfpage{9682}
(\byear{1990})
\end{barticle}
\endbibitem

\bibitem{maginn2019best}
\begin{barticle}
\bauthor{\bsnm{Maginn}, \binits{E.J.}},
\bauthor{\bsnm{Messerly}, \binits{R.A.}},
\bauthor{\bsnm{Carlson}, \binits{D.J.}},
\bauthor{\bsnm{Roe}, \binits{D.R.}},
\bauthor{\bsnm{Elliot}, \binits{J.R.}}:
\batitle{Best practices for computing transport properties 1. self-diffusivity
  and viscosity from equilibrium molecular dynamics [article v1. 0]}.
\bjtitle{Living Journal of Computational Molecular Science}
\bvolume{1}(\bissue{1}),
\bfpage{6324}--\blpage{6324}
(\byear{2019})
\end{barticle}
\endbibitem

\bibitem{chen2013melting}
\begin{barticle}
\bauthor{\bsnm{Chen}, \binits{M.}},
\bauthor{\bsnm{Hung}, \binits{L.}},
\bauthor{\bsnm{Huang}, \binits{C.}},
\bauthor{\bsnm{Xia}, \binits{J.}},
\bauthor{\bsnm{Carter}, \binits{E.A.}}:
\batitle{The melting point of lithium: an orbital-free first-principles
  molecular dynamics study}.
\bjtitle{Molecular Physics}
\bvolume{111}(\bissue{22-23}),
\bfpage{3448}--\blpage{3456}
(\byear{2013})
\end{barticle}
\endbibitem

\bibitem{neumann2011selfdiffusion}
\begin{bbook}
\bauthor{\bsnm{Neumann}, \binits{G.}},
\bauthor{\bsnm{Tuijn}, \binits{C.}}:
\bbtitle{Self-diffusion and Impurity Diffusion in Pure Metals: Handbook of
  Experimental Data}.
\bpublisher{Elsevier}, \blocation{???}
(\byear{2011})
\end{bbook}
\endbibitem

\bibitem{gd40}
\begin{botherref}
\oauthor{\bsnm{Yu}, \binits{T.}},
\oauthor{\bsnm{Zhu}, \binits{H.}}:
Hyper-parameter optimization: A review of algorithms and applications.
arXiv preprint arXiv:2003.05689
(2020)
\end{botherref}
\endbibitem

\bibitem{gd41}
\begin{barticle}
\bauthor{\bsnm{Dhaliwal}, \binits{G.}},
\bauthor{\bsnm{Nair}, \binits{P.B.}},
\bauthor{\bsnm{Singh}, \binits{C.V.}}:
\batitle{Uncertainty analysis and estimation of robust airebo parameters for
  graphene}.
\bjtitle{Carbon}
\bvolume{142},
\bfpage{300}--\blpage{310}
(\byear{2019})
\end{barticle}
\endbibitem

\bibitem{gd42}
\begin{barticle}
\bauthor{\bsnm{Dhaliwal}, \binits{G.}},
\bauthor{\bsnm{Nair}, \binits{P.B.}},
\bauthor{\bsnm{Singh}, \binits{C.V.}}:
\batitle{Uncertainty and sensitivity analysis of mechanical and thermal
  properties computed through embedded atom method potential}.
\bjtitle{Computational Materials Science}
\bvolume{166},
\bfpage{30}--\blpage{41}
(\byear{2019})
\end{barticle}
\endbibitem

\end{thebibliography}
\end{document}